\documentclass[twocolumn, amssymb,amsmath, nobibnotes, aps,pra, reprint]{revtex4-1}
\usepackage{graphicx}
\usepackage{dcolumn}
\usepackage{bm}
\usepackage{flushend}
\usepackage{balance}
\usepackage{lipsum}
\usepackage{mathtools}
\usepackage{cuted}

\usepackage[caption=false]{subfig}
\usepackage{float}
\usepackage{silence}
\newcommand{\be}{\begin{equation}}

\newcommand{\ee}{\end{equation}}
\newcommand{\ben}{\begin{eqnarray}}
\newcommand{\een}{\end{eqnarray}}
\newcommand{\bes}{\begin{subequations}}
\newcommand{\ees}{\end{subequations}}
\newcommand{\bF}{\begin{figure}}
\newcommand{\eF}{\end{figure}}

\def\tr{ {\rm{Tr }}\,}

\newcommand{\kt}{\rangle}
\newcommand{\br}{\langle}
\newcommand{\ket}[1]{\left|#1\right\rangle}
\newcommand{\bra}[1]{\left\langle#1\right|}

\newcommand{\avg}[1]{\left\langle #1 \right\rangle}

\newcommand{\mbf}[1]{\mathbf{#1}}
\usepackage{lineno}
\usepackage[colorlinks,citecolor=red,urlcolor=blue,bookmarks=false,hypertexnames=true]{hyperref} 

\modulolinenumbers[5]

\begin{document}

\title{Effect of chaos on information gain in quantum tomography}%

\author{Abinash Sahu}%
\email{ommabinash@physics.iitm.ac.in}
\affiliation{Department of Physics, Indian Institute of Technology Madras, Chennai, India, 600036}

\author{Sreeram PG}
\email{sreerampg@outlook.com}
\affiliation{Department of Physics, Indian Institute of Technology Madras, Chennai, India, 600036}

\author{Vaibhav Madhok}
\email{madhok@physics.iitm.ac.in}
\affiliation{Department of Physics, Indian Institute of Technology Madras, Chennai, India, 600036}

\begin{abstract}

Does chaos in the dynamics enable information gain in quantum tomography or impede it? 
We address this question by considering continuous measurement tomography in which the measurement record is obtained as a
sequence of expectation values of a Hermitian observable evolving under the repeated application of
the Floquet map of the quantum kicked top. For a given dynamics and Hermitian observables, we observe completely opposite behavior in the tomography of well-localized spin coherent states compared to random states. As the chaos in the dynamics increases, the reconstruction fidelity of spin coherent states decreases. This contrasts with the previous results connecting information gain in tomography of random states with the degree of chaos in the dynamics that drives the system.  
The rate of information gain and hence the fidelity obtained in tomography depends not only on the degree of chaos in the dynamics and to what extent it causes the initial observable to spread in various directions of the operator space but, more importantly, how well these directions are aligned with the density matrix to be estimated. Our study also gives an operational interpretation for operator spreading in terms of fidelity gain in an actual quantum information tomography protocol.

\end{abstract}

\maketitle

\section{Introduction}
The rapid divergence of neighboring classical trajectories with time, often described as exponential sensitivity to initial conditions, is the hallmark of deterministic chaos in classical mechanics. However, this kind of divergence between two possible initial states cannot occur quantum mechanically because of the necessity of preserving the inner product due to the linearity of Schr\"{o}dinger's equation. One of the primary goals in the field of quantum chaos is to search for the signatures of chaos in quantum systems and their consequences in quantum information processing, statistical mechanics, in foundational areas like quantum-to-classical transition, and the rate of decoherence under chaotic dynamics. Various signatures of chaos have been discovered. Starting from the  behavior of the spectral statistics of the generating Hamiltonian ~\cite{haake1991quantum} to  dynamical signatures of chaos like hypersensitivity of system dynamics to  perturbation~\cite{peres1984stability,schack1996information} and the dynamical generation of quantum correlations, such as quantum entanglement~\cite{miller1999signatures,bandyopadhyay2002testing,wang2004entanglement,trail2008entanglement,furuya1998quantum,lakshminarayan2001entangling,seshadri2018tripartite}, and quantum discord~\cite{madhok2015signatures,madhok2018quantum}. Recently, out-of-time ordered correlators have also been used to probe quantum chaos~\cite{maldacena2016bound,swingle2016measuring,
hashimoto2017out,kukuljan2017weak,swingle2018unscrambling,wang2021quantum,sreeram2021out,varikuti2022out}. The signatures of chaos are not only explored in the semiclassical limit but also in the deep quantum regime~\cite{neill2016ergodic,fortes2020signatures,sreeram2021out}. In this work, we investigate the role of chaos in the rate of information gain in tomography, which has been shown to be a quantum signature of chaos in ~\cite{madhok2014information}. {We aim to give a complete picture elucidating the role of dynamics and prior information about the state relative to the operators measured and the role of noise in tomography. }

 Tomography of quantum states is essential for quantum information processing tasks like quantum computation, quantum cryptography, quantum simulations, and quantum control. Estimation of quantum states is a highly nontrivial problem because of fundamental restrictions posed by Heisenberg's uncertainty principle and no-cloning theorem  \cite{wootters1982single}. Different protocols have carried out tomography in many systems~\cite{paris2004quantum,d2003quantum}. State reconstruction uses the statistics of measurement records on an ensemble of identical systems in order to make the best estimate of the actual state $\rho_0$. An informationally complete set of measurement records is required for high fidelity tomography. Inverting these records, in principle should give an estimate of the state. The traditional way has been using projective measurements to extract the information. However, such protocols are resource-intensive since strong measurements destroy the state. To get good fidelity reconstruction, one would then require infinitely many copies of the system. Weak measurement alternatives have been explored in the literature~\cite{lundeen2011direct, wu2013state,hofmann2010complete, shojaee2018optimal,silberfarb2005quantum,smith2006efficient,chaudhury2009quantum,merkel2010random,smith2004continuous}. Weak measurements help in reducing the number of copies of the system required for the process since they cause minimal disturbance to the system. However, the amount of information gained per measurement is bound to be low in this type of measurement \cite{busch2009no}. In this article, we are interested in continuous weak measurement tomography~\cite{silberfarb2005quantum,smith2006efficient,chaudhury2009quantum,merkel2010random,smith2004continuous}. A time series of operators is generated by a single parameter unitary~\cite{merkel2010random,PhysRevA.104.032404} or the Floquet map of the quantum kicked top~\cite{madhok2014information} in the Heisenberg picture and measurement record is obtained. 

The central focus of our work is the following question: How reconstructing quantum states is related to the nature of dynamics employed in the tomography process? At first, the connection between chaos and state reconstruction seems distant. Chaos is about the inability to predict the long-term behavior of a dynamical system, while tomography involves information acquisition. However, the flip side of this uncertainty and unpredictability of chaotic dynamics is information. If everything is known about a trajectory, for example, a periodic orbit, we gain no new information.
Classically, as one tracks a chaotic trajectory, one gains information at a rate proportional to the magnitude of chaos in the system. This rate is more formally described as the Kolmogorov-Sinai (KS) entropy and is equal to the sum of positive Lyapunov exponents of the system~\cite{pesin1977characteristic}. One might ask what this information is about? The answer is  \textit{initial conditions}. One obtains information on increasingly finer scales about the system's initial conditions. In quantum mechanics, this is precisely the goal of tomography. As one follows the archive of the measurement record in a tomography experiment, one gains information about the initial random quantum state. An intriguing question in the quantum case is whether or not the rate of information gain is related to the degree of chaos in the dynamics. 
There seems an fascinating and provocative connection between tomography and chaos as demonstrated in ~\cite{madhok2014information}. While  \cite{madhok2014information} considered quantum tomography for random states, we find that state reconstruction for localized wave packets remarkably shows opposite behavior! We show that the rate of information gain is a function of dynamics and the initial state as well as the relationship between time evolved operators and the initial state. 

The remainder of this article is organized as follows. In the next section, we provide background information on the concepts and tools we use in this work. Section III gives an overview of continuous weak measurement tomography. In section IV, the heart of the manuscript, we explore the relationship between tomography and dynamics for localized spin coherent states and contrast them with that for random states. Finally, we conclude by discussing our findings in the last section.

\section{Background}
\subsection{Quantum kicked top}
Quantum kicked top is a time-dependent periodic system governed by the Hamiltonian ~\cite{haake1987classical,haake1991quantum,chaudhury2009quantum}
\begin{equation}
H=\hbar\alpha J_x+\hbar\frac{\lambda }{2j\tau}J^2_z\ \sum_{n}\delta(t-n\tau) 
\label{H1}.
\end{equation}
Here $J_x$, $J_y$, and $J_z$ are the components of the angular momentum operator $\bf J$. The first term in the Hamiltonian $H$ describes a linear precession of $\bf J$ around the $x$-direction by an angle $\alpha$. Each kick is a nonlinear rotation about $z$-direction in a periodic time interval of $\tau$, as given in the second term of the Hamiltonian. The strength of this nonlinear rotation is $\lambda$ and is also the chaoticity parameter. The delta kick allows us to express the Floquet map as a sequence of operations given by 
\begin{equation}
 U_\tau=\mathrm{exp}\Big(-i\frac{\lambda}{2j\tau}J^2_z\Big)\ \mathrm{exp}(-i\alpha J_x).
\end{equation}
Thus, the time evolution unitary for time $t=n\tau$, $n=0, 1, 2, ...$ is $U(n\tau)=U^n_\tau$. The Heisenberg evolution of an operator generates a sequence of operators $\mathcal{O}_n=U^{\dagger n}\mathcal{O}U^{n}$. \\

The classical behavior of this map can be seen by expressing the Heisenberg equations of motion for the angular momentum operators $J_x$, $J_y$, and $J_z$ and then taking the limit $j\rightarrow \infty$. The resulting equations describe the motion of an angular momentum vector on the surface of a sphere. The dynamics can be realized as a linear precession by an angle $\alpha$ about the $x-$axis, followed by a nonlinear precession about the $z-$axis. The absence of enough constants of motion in the system leads to chaotic dynamics. For our current work, we fix $\alpha=\pi/2$ and choose $\lambda$ as the chaoticity parameter. The classical dynamics change from highly regular to fully chaotic as we vary $\lambda$ from $0$ to $7$.

\subsection{Spin coherent states}

To address the problem, we will explore the reconstruction of random states and coherent states which are on two extremes as far as localization in phase space is concerned. Being the closest analog of classical minimum uncertainty wave packets, coherent states  \cite{bohr1920seri} are particular quantum states of a quantum harmonic oscillator that, despite the quantum mechanical uncertainty in position and momentum, follow classical-like dynamics. Similarly, spin coherent states are minimum uncertainty wave packets that satisfy the Heisenberg uncertainty principle for angular momentum operators.

Spin coherent states are highly localized and serve as the closest analog for points in a classical phase space. The spin coherent states point in a particular direction to the extent allowed by the angular momentum commutation relation. For a given point $(\theta,\phi)$ in the classical phase space, the spin coherent state is defined as~\cite{arecchi1972atomic,radcliffe1971some}

\begin{equation}
\ket{\theta,\phi}=(1+|\mu|^2)^{-j}e^{\mu{J_-}}\ket{j,j}\equiv \ket{\mu}                                                                                                                                                                                                                                                                                                                                                                                                                                                                                                                                                                                                                                                                                                             \end{equation}
where $\mu=e^{i\phi}\tan\frac{\theta}{2}$; $0\leq\theta\leq\pi$, $0\leq\phi\leq2\pi$, and the spin lowering operator $J_-=J_x-iJ_y$. The spin coherent state $\ket{j,j}$ is one of the eigenbasis of $J^{2}$ and $J_z$ from the set $\{\ket{j,m}\}, m=-j,-j+1,...,j$. Other directed angular momentum states can be generated by rotating the state $\ket{j,j}$ as
\begin{equation}
\ket{\theta,\phi}=\mathrm{exp}\Big(i\theta(J_x{\sin\phi}-J_y{\cos\phi})\Big)\ket{j,j}. 
\end{equation}
The uncertainty in $J$ for the state $\ket{\theta,\phi}$ is  
\begin{equation}
 (\avg{J^2}-{\avg{J}}^2)/j^2=1/j
\end{equation}
Thus, the uncertainty goes to zero as the $j$ value becomes very large, and the spin coherent states are highly localized at the point $(\theta,\phi)$ in the phase space.
\subsection{Husimi entropy}
The Husimi $\mathcal{Q}$-function of a density matrix $\rho$ is a quasiprobability distribution in phase space. Husimi $\mathcal{Q}$-function is defined as \cite{husimi1940some}
\begin{equation}
 \mathcal{Q}_{\rho}(\theta,\phi)=\br\theta,\phi|\rho|\theta,\phi\kt.
\end{equation}
A notion of entropy can be associated with any density matrix through  Husimi $\mathcal{Q}$-function, called the Husimi entropy (also known as Wehrl entropy) \cite{wehrl1979relation},
\begin{equation}
 S_\rho=-\frac{2j+1}{4\pi}\int_{\Omega}d\Omega\ \mathcal{Q}_{\rho}ln[\mathcal{Q_\rho}].
\end{equation}
where $\Omega= \lbrace \theta, \phi \rbrace,\ 0 \leq \theta \leq \pi,\ 0\leq \phi\leq 2\pi$ . To treat both observables and  density operators on equal footing, we  determine the Husimi entropy for an operator after doing some regularization as follows. We construct a positive operator from an observable by retaining its eigenvectors and taking the modulus of its eigenvalues.  To normalize this operator, we divide by its trace. Now we  can calculate the Husimi entropy and  analyze the localization of the operator in the phase space. For regularized Hermitian observables the Husimi function is the expectation value with respect to spin coherent state $\ket{\theta,\phi}$. Thus, for a regularized operator $\mathcal{O}$ the Husimi function is 
\begin{equation}
 \mathcal{Q}_{\mathcal{O}}(\theta,\phi)=\br\theta,\phi|\mathcal{O}|\theta,\phi\kt,
 \label{hs1}
\end{equation}
 and the Husimi entropy is given by
\begin{equation}
 S_\mathcal{O}=-\frac{2j+1}{4\pi}\int_{\Omega}d\Omega\ \mathcal{Q}_{\mathcal{O}}ln[\mathcal{Q_\mathcal{O}}].
 \label{hs2}
\end{equation}

\section{Continuous measurement tomography}
We are given an ensemble of $N$ identical systems $\rho^{\otimes N}_0$ and they undergo separable time evolution by a unitary $U(t)$. A probe is coupled to the ensemble of states that will generate the measurement record by performing weak continuous measurement of the observable $\mathcal{O}$. We use the Heisenberg picture, and  the operator that is measured at time $t$ is 
\begin{equation}
 \mathcal{O}(t)=U^{\dag}(t)\mathcal{O}U(t)
\end{equation}

The positive operator valued measurement (POVM) elements for measurement outcomes $X(t)$ at time $t$ are~\cite{silberfarb2005quantum,madhok2014information} 
\begin{equation}
 E_{X(t)}=\frac {1}{\sqrt{2\pi\sigma^2}}\mathrm{exp}\ \Big\{-\frac{1}{2\sigma^2}[X(t)-\mathcal{O}(t)]^2\Big\},
\end{equation}
The standard deviation $\sigma$ in the POVM elements is due to the shot noise of the probe. When the randomness of the measurement outcomes is dominated by the quantum noise in the probe rather than the measurement uncertainty, i.e., the projection noise, quantum backaction is negligible, and the state remains approximately separable. Thus the measurement records can be approximated to be 

\begin{equation}
 M(t)=X(t)/N=Tr[\mathcal{O}(t)\rho_0]+W(t)
 \label{tom_noise}
\end{equation}
where $W(t)$ is a Gaussian white noise with spread $\sigma/N$.

The density matrix of any arbitrary state having Hilbert space dimension $d$ can be expressed in the orthonormal basis of $d^2-1$ traceless and Hermitian operators $\{E_\alpha\},$ and the state lies on the generalized Bloch sphere parametrized by the Bloch vector $\bf r$. Thus the density matrix can be represented as 
\begin{equation}
\rho_0=I/d+\Sigma^{d^2-1}_{\alpha=1}\ r_\alpha E_\alpha,  
\label{dens_matrix}
\end{equation}
where $$\Sigma^{d^2-1}_{\alpha=1}\ r_\alpha^2 = 1 - 1/d  $$

We consider the measurement records at discrete times 
\begin{equation}
 M_n=M(t_n)=N\sum_{\alpha}r_\alpha \tr[\mathcal{O}_{n}E_\alpha]+W_n,
\end{equation}
 where $\mathcal{O}_n=U^{\dagger n}\mathcal{O}U^{n}$.
Thus, in the negligible backaction limit, the probability distribution associated with measurement history $\bf M$ for a given state vector $\bf r$ is \cite{silberfarb2005quantum,smith2006efficient} is
\begin{equation} 
\begin{split}
p({\bf M|r}) & \varpropto \mathrm{exp}\ \Big\{-\frac {N^2}{2\sigma^2}\sum_{i}[M_i-\sum_{\alpha}\tilde{\mathcal{O}}_{i\alpha}r_\alpha]^2\Big\}
\\
& \varpropto \mathrm{exp}\ \Big\{-{\frac {N^2}{2\sigma^2}\sum_{\alpha,\beta}({\bf r-r_{ML}})_\alpha\ C^{-1}_{\alpha\beta}\ ({\bf r-r_{ML}})_\beta\Big\}},
\label{pdf_uniform_prior}
\end{split}
\end{equation}
where $\tilde{\mathcal{O}}_{n\alpha}=\tr[\mathcal{O}_{n}E_\alpha]$ and ${\bf C^{-1}}=\tilde{\mathcal{O}}^T\tilde{\mathcal{O}}$ is the inverse of the covariance matrix. Given the measurement record and the knowledge of the dynamics, one can invert this measurement record to get an estimate of the parameters characterizing the unknown quantum state in Eq. (\ref{dens_matrix}).
The least-square fit of the Gaussian distribution in the parameter space is the maximum-likelihood (ML) estimation of the Bloch vector, ${\bf r}_{ML}=\bf C\tilde{\mathcal{O}}^{T}M.$ The measurement record is informationally complete if the covariance matrix is full rank. If the covariance matrix is not full rank, the inverse of the covariance matrix is replaced by Moore-Penrose pseudo inverse~\cite{ben2003generalized}, inverting over the subspace where the covariance matrix has support. The eigenvalues of the $\bf C^{-1}$ determine the relative signal to noise ratio with which different observables have been measured. The estimated Bloch vector ${\bf r}_{ML}$ may not represent a physical density matrix with non-negative eigenvalues because of the noise present (having a finite signal to noise ratio). Therefore we impose the constraint of positive semidefiniteness~\cite{baldwin2016strictly}  on the reconstructed density matrix and obtain the  physical state closest to the maximum-likelihood estimate.

To do this, we employ a convex optimization~\cite{vandenberghe1996semidefinite} procedure where the final estimate of the Bloch vector $\bf \bar{r}$ is obtained by minimizing the argument 
\begin{equation}
 ||{\bf r}_{ML}-{\bf \bar{r}}||^2=({\bf r}_{ML}-{\bf \bar{r}})^T{\bf C}^{-1}({\bf r}_{ML}-\bf \bar{r})
\end{equation}
subject to the constraint $$I/d+\Sigma^{d^2-1}_{\alpha=1}\ \bar{r}_\alpha E_\alpha\geq0.$$ {The positivity constraint plays a crucial role in compressed sensing  tomography  as well.  Any optimization heuristic with positivity constraint is effectively a compressed sensing protocol, provided that the measurements are within the special class associated with compressed sensing~\cite{kalev2015quantum}. }

In this article, we use the periodic application of a Floquet map $U_\tau$ for simplicity, and the unitary at $n^{\mathrm{th}}$ time step is $U(n\tau)=U^n_\tau$. The measurement record generated by such periodic evolution is not informationally complete, and it leaves out a subspace of dimension $\geq d-2,$  out of $d^2-1$ dimensional operator space. We employ a well-studied kicked top model \cite{haake1987classical,haake1991quantum,chaudhury2009quantum} described by the Floquet map $U_\tau=e^{-i\lambda J^2_z/2J}e^{-i\alpha J_x}$ as the unitary. 

\begin{figure*}[htbp]
 \centering
  \subfloat[\label{fig:fid1a}]{%
    \includegraphics[width=8.5cm, height=6.1cm]{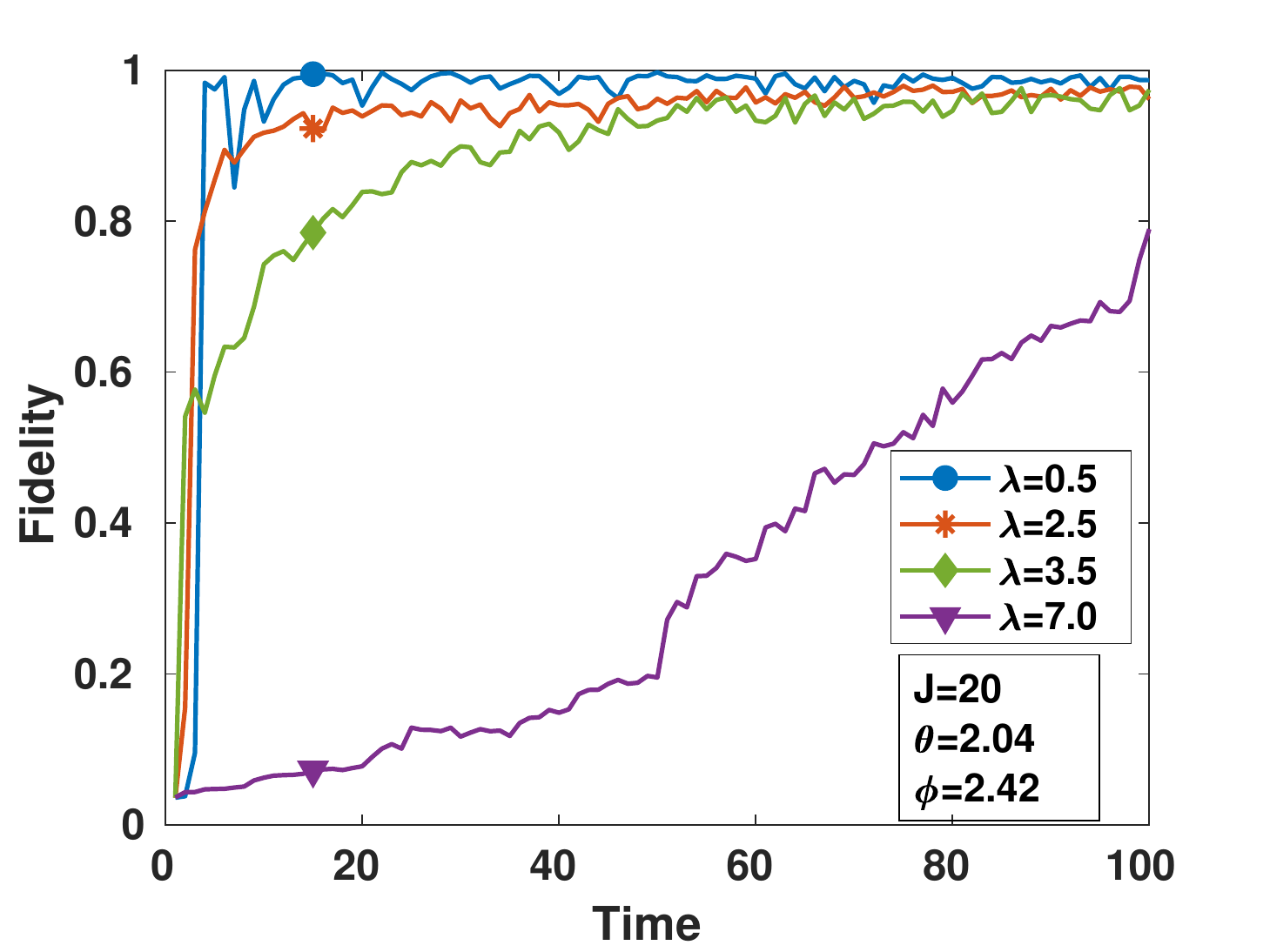}}\hfill
  \subfloat[\label{fig:fid1b}]{%
    \includegraphics[width=8.5cm, height=6.1cm]{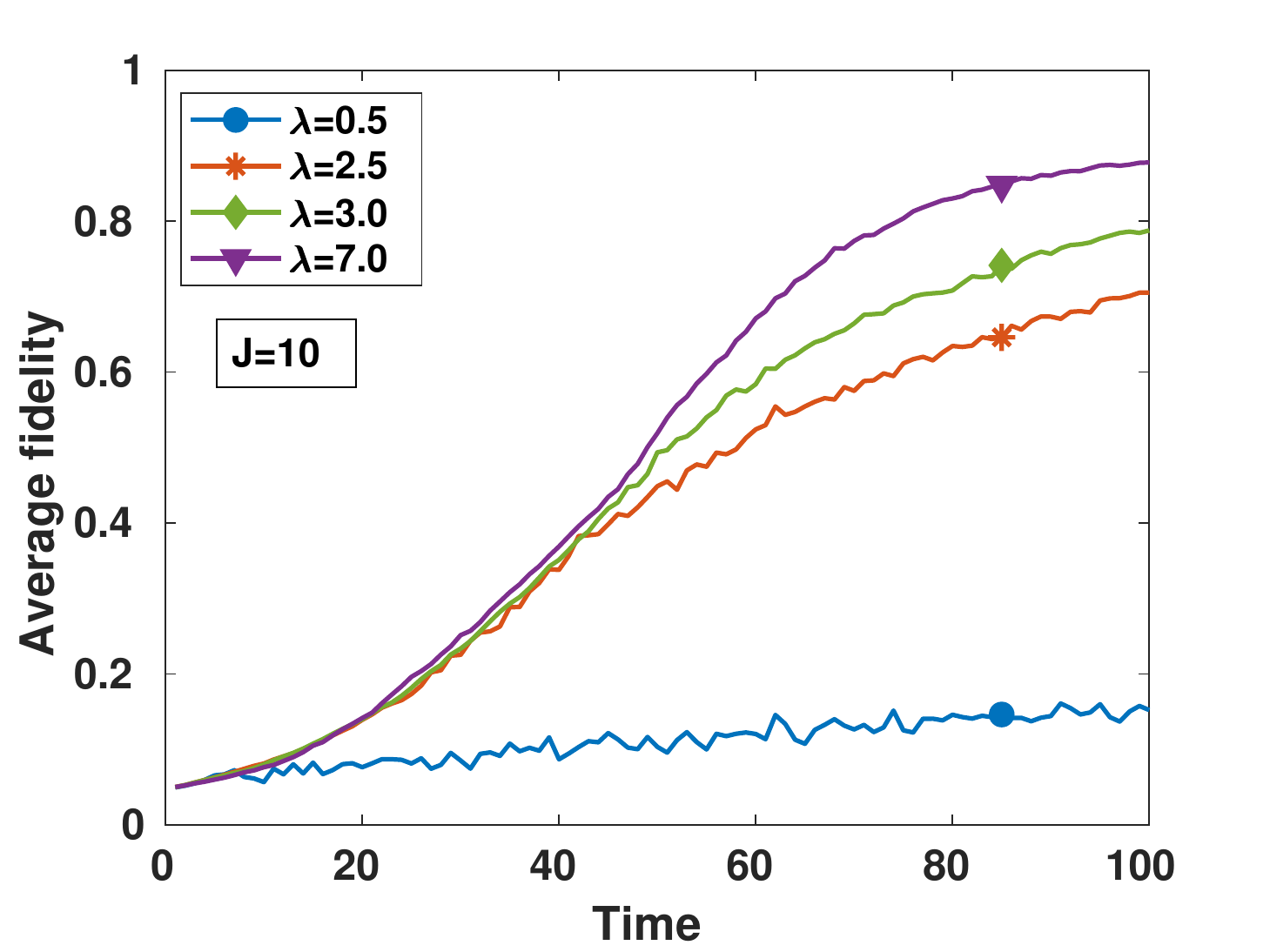}}\hfill
    \hfill
  \subfloat[\label{fig:fid1c}]{%
    \includegraphics[width=8.5cm, height=6.1cm]{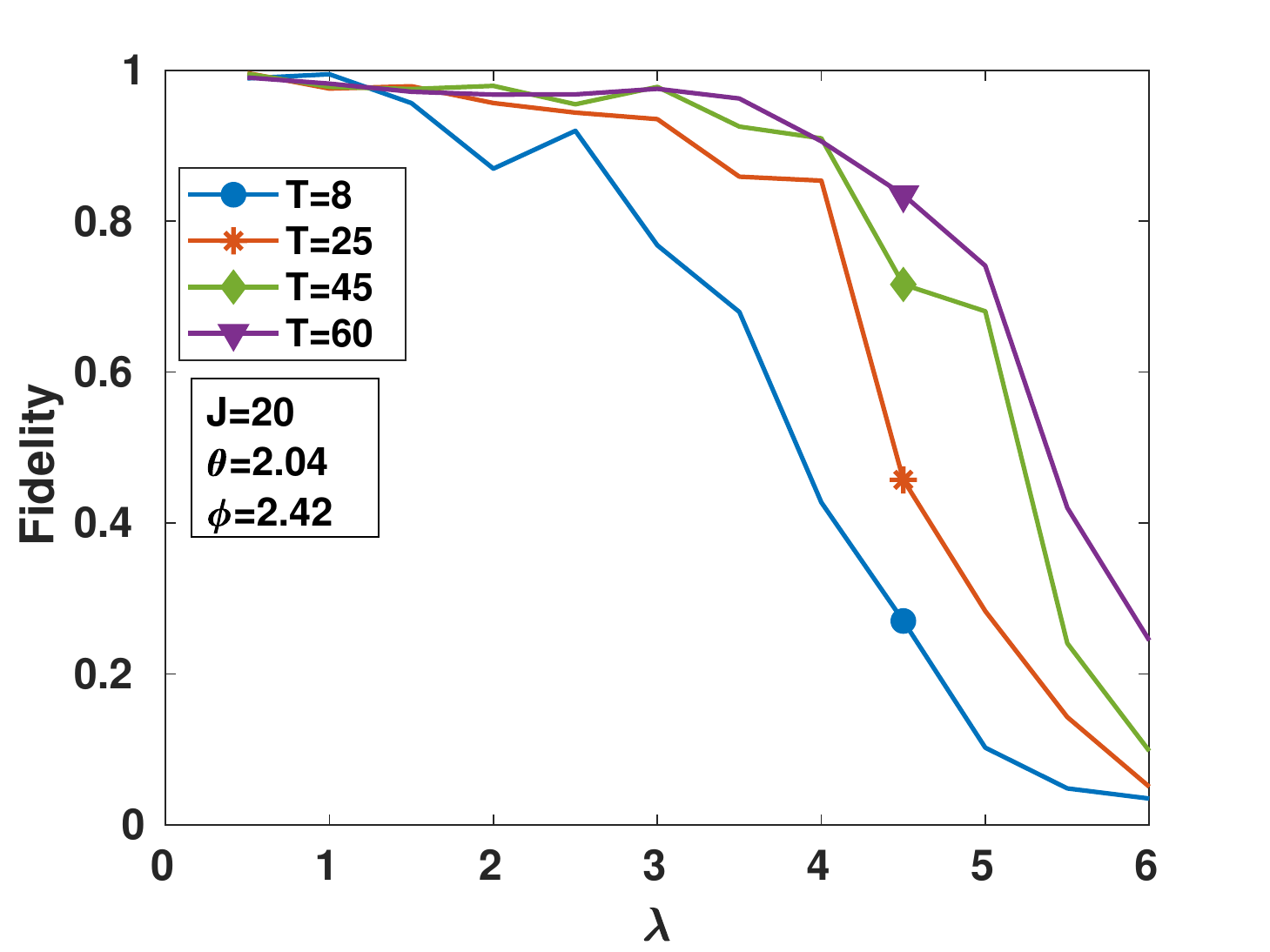}}\hfill
    \hfill
  \subfloat[\label{fig:fid1d}]{%
    \includegraphics[width=8.5cm, height=6.1cm]{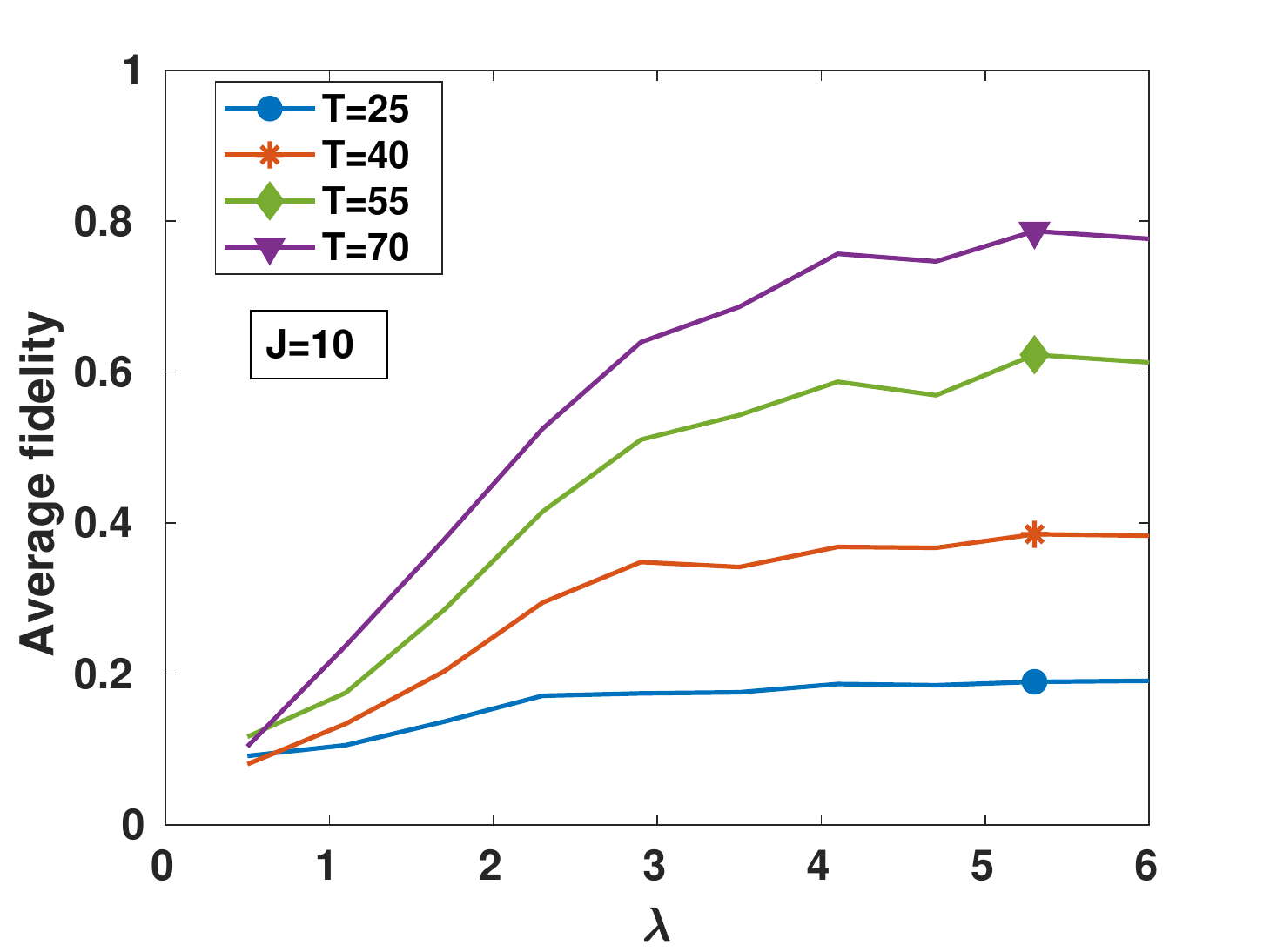}}
 \caption{Column-wise comparison showing the contrasting behaviour of reconstruction fidelity for spin coherent state and random states. (a and b) show fidelity as a function of time for different chaoticity parameters. (c and d) show fidelity as a function of chaoticity at different time steps of the tomography process. The fidelity of spin coherent state with $\theta=2.04$ and $\phi=2.42$ (a random choice from the phase space) decreases with the increase in the values of the chaoticity parameter $\lambda$, whereas average fidelity of the random states increases with increase in $\lambda.$ }
 \label{fig:fid1}
\end{figure*}

\section{Quantum chaos and Tomography: spin coherent states vs. random states}
In this section, we come to the central question we ask. What is the effect of the degree of chaos on the tomography of states? 
{ For our analysis, we study the dynamics of quantum kicked top for angular momentum $j=20$ for spin coherent states and $j=10$ for random states.} For the tomography, we consider the initial observable as $\mathcal{O}=J_y$, and the subsequent observables whose expectation values we measure are acquired by evolving under the Floquet map of the quantum kicked top. The fidelity of the reconstructed state $\bar{\rho}$ is determined relative to the actual state $\ket{\psi_0},$ $\mathcal{F}=\bra{\psi_0}\bar{\rho}\ket{\psi_0}$ as a function of time.

We  discover interesting, contrasting, and counterintuitive effects of chaos in the tomography fidelities depending on whether the states involved are random states spread across the phase space or localized coherent states. Figure \ref{fig:fid1a} and Fig. \ref{fig:fid1b} show the reconstruction fidelities as a function of time for coherent states and random states respectively with different degrees of chaos. A common observation in both cases is that as time increases, the fidelity rises. This is because  of more measurements and information gain with time. However, with chaoticity, the coherent and random states show opposite behavior. It is evident from Fig. \ref{fig:fid1a} that for spin coherent states the fidelity decreases with the increase in the level of chaos, which is in contrast to the nature of random states~\cite{madhok2014information}, as shown in Fig. \ref{fig:fid1b}. {This is made more clear in Fig. \ref{fig:fid1c} and  \ref{fig:fid1d}, where we plot fidelity against chaoticity at different instances of time.} We set out to investigate this distinctive behavior with respect to chaoticity.

First, we ask the following question: what constitutes information gain in tomography? More precisely, the rate of information gain in tomography? It is important to make this crucial distinction between information gain and  its acquisition rate for the following reason. In the limit of vanishing shot noise in Eq. (\ref{tom_noise}) and assuming an informationally complete measurement record, we can reconstruct the quantum state with unit fidelity irrespective of the dynamics involved. This is because we are able to determine the components of the $d^2-1$ dimensional generalized Bloch vector completely from such a noiseless and informationally complete measurement record. This can be seen, for example, in the case of quantum tomography for a single qubit on the usual Bloch vector on the 2-sphere. Here we need three expectation values in the direction of the Pauli matrices to determine the components of this vector and completely specify the state.

\begin{figure}[htb]
 \centering
 \includegraphics[width=8.5cm, height=6.1cm]{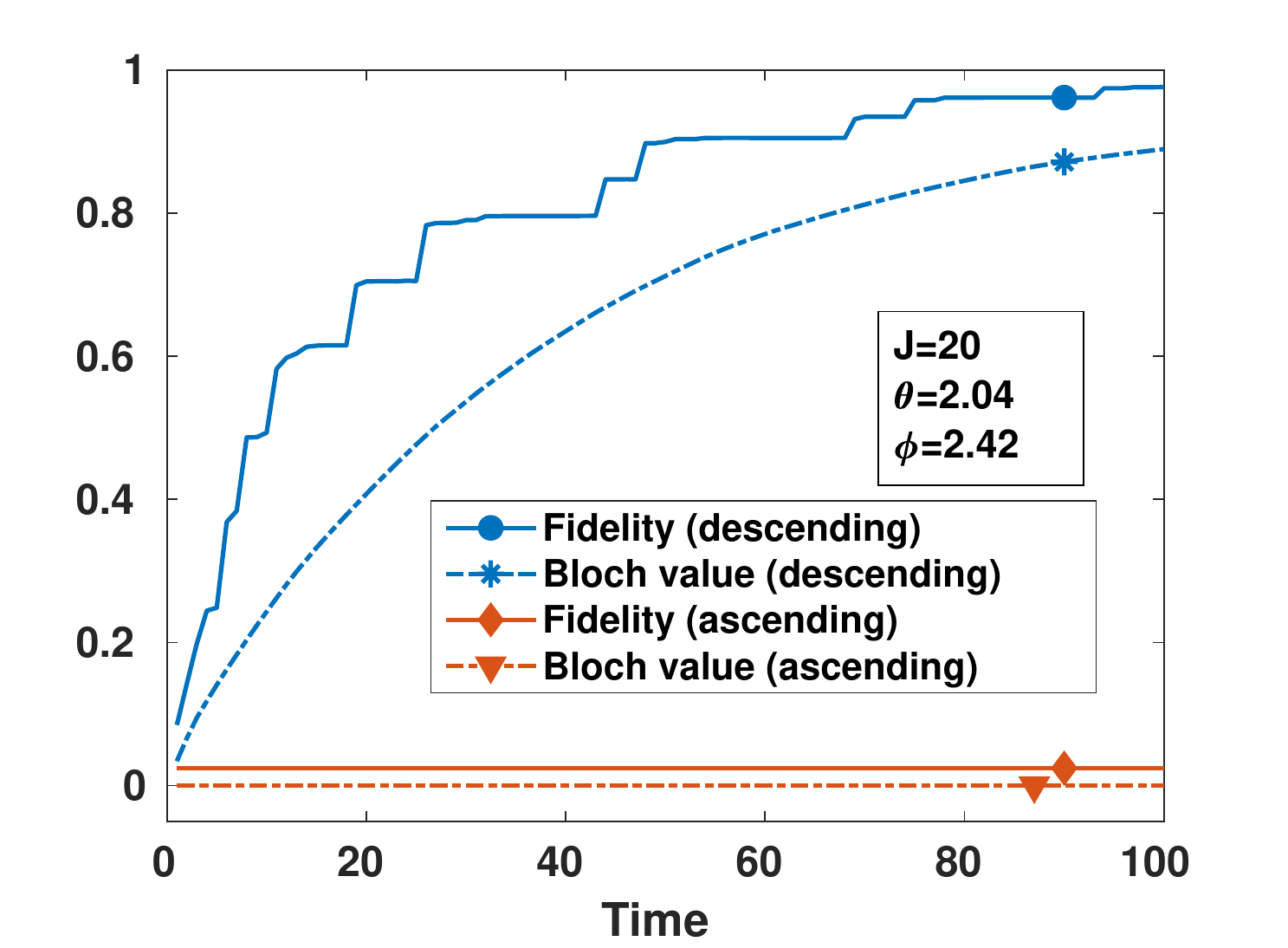} 
  
 \caption{Information gain (Bloch values)  with ordered Bloch vector components  (i.e. that corresponds to the Bloch vector components, $r_{\alpha} = \tr[\rho_0 E_{\alpha}]$  in the descending and ascending order of magnitude), and fidelity in the limit of vanishing shot noise. The Bloch value  at time $k$ refers to the quantity  $r_{i}^2 +  r_{j}^2+ ...+ r_{k}^2$, where $\{r_i,r_j,...,r_k\}$ is the ordered set of Bloch vector components as described above.}
 \label{fig:gell}
\end{figure}

However, even in the case of vanishing shot noise (that gives us the maximal signal to noise ratio), the order in which we measure various operators matters as far as the rate of information gain is concerned. For example, let us consider the density matrix as a vector and express it as Eq. (\ref{dens_matrix}), we ask the following question: What is the order in which one should measure various $E_\alpha$'s to get the most rapid information gain about the unknown state? It is easy to see that the order of \{$E_\alpha$\}, that corresponds to the Bloch vector components \{$r_\alpha$\} in the descending order of magnitude gives the maximum rate of information gain. Figure \ref{fig:gell} shows the effect of the ordering of \{$E_\alpha$\} on the fidelity and information gain in tomography.

The above discussion helps us qualitatively understand why spin coherent states do worse in reconstruction as one increases the chaos in the dynamics. Chaos scrambles and delocalizes the operators such that the subsequent operators generated in the Heisenberg picture have less support over the density matrix, which hinders rapid information gain. The fidelities obtained are a function of the dynamics and the degree to which the operators generated yield information about the Bloch vector components.
However, we need to elucidate this intuition with a more concrete analysis.

The probability distribution of observing a measurement record $\bf M$ given an initial state $\rho_0,$  the dynamics $\mathcal{L}$ (that involves application of unitaries),  and the measurement process $\mathcal{M}$ (the choice of operators $\mathcal{O}$ to be measured) is $p({\bf M|\rho_0, \mathcal{L}, \mathcal{M}})$. Thus the probability of reconstructing the state $\rho_0$  is

\begin{equation}
p({\bf \rho_0|M, \mathcal{L}, \mathcal{M}}) = A \: p({\bf M|\rho_0, \mathcal{L}, \mathcal{M}})\: p(\rho_0|\mathcal{L}, \mathcal{M})\: p(\mathcal{L}, \mathcal{M}).
\label{tom_eq}
\end{equation}
Here $A$ is a normalization constant, $p(\rho_0|\mathcal{L}, \mathcal{M})$ is the  posterior probability distribution conditioned upon the knowledge of the dynamics and the measurement operators. In the limit of zero noise, and given measurement observables \{$E_\alpha$\}, this conditional probability is constantly updated and eventually becomes a product of Dirac-delta functions, each of them specifying a particular Bloch vector component,  once we obtain an informationally complete measurement record. The term, $ p(\mathcal{L}, \mathcal{M})$, in the above expression, is the prior information about the choice of dynamics and measurement operators and can be absorbed in the constant.
Equation (\ref{tom_eq}) is illuminating as it separates 
the probability of estimation into a product of two terms (up to a constant). The first term $p({\bf M|\rho_0, \mathcal{L}, \mathcal{M}})$, which is identical to Eq. (\ref{pdf_uniform_prior}), contains the errors due to shot noise and quantifies the signal to noise ratio in various directions in the operator space \textit{independent} of the state to be estimated.  Therefore, this term estimates the information gained, given a density matrix, in different directions in the operator space. The second term quantifies how likely this particular density matrix is to be the actual unknown initial state. This gives a constant factor for random states as there is no correlation between the measurement observables and the initial state chosen randomly.

{However, for spin coherent state tomography,  the  term $p(\rho_0|\mathcal{L}, \mathcal{M})$ becomes crucial, as we see in the discussion below. Let us look at a measure of information gain that is oblivious to the choice of initial state and reordering of measurement operators.} We can  quantify the correlation between system dynamics and information gain in quantum tomography by calculating the Fisher information associated with the measurement process. {For the case of random states, this measure perfectly characterizes the effect of chaos on tomography \cite{PhysRevA.104.032404,madhok2014information}.} Quantum tomography is equivalent to ``parameter estimation'',  i.e., estimation of the Bloch vector components that define the density matrix $\rho_0$. The Fisher information quantifies how well our estimator can predict these parameters from the data, regardless of the state.

\begin{figure}[htb]
 \centering
 \includegraphics[width=8.5cm, height=6.1cm]{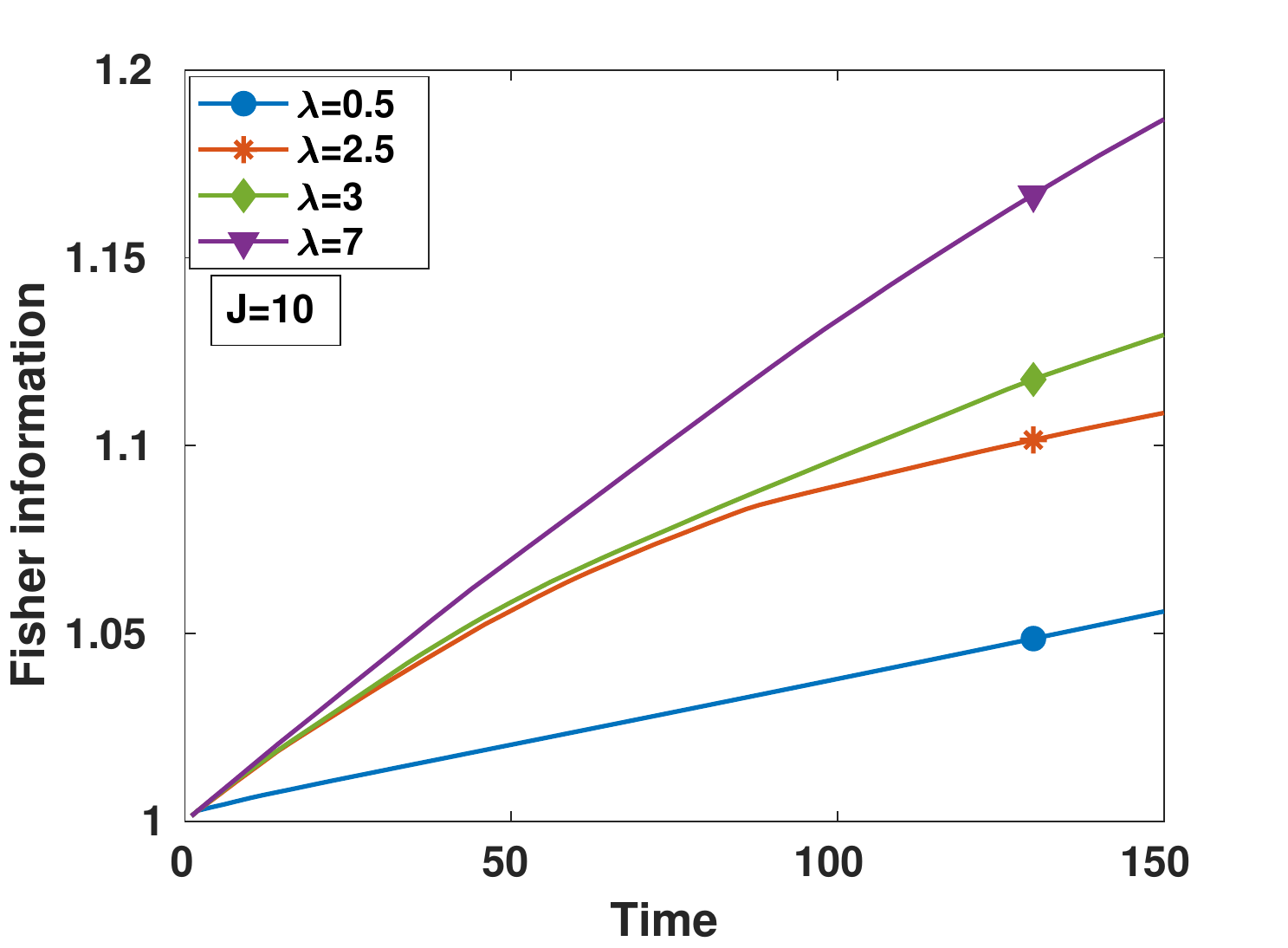} 
  
 \caption{The Fisher information of the parameter estimation in tomography as a function of time for different degrees of chaos.}
 \label{fig:FIrnd}
\end{figure}

 The  Hilbert-Schmidt distance between the true and estimated state in quantum tomography, averaged over many runs of the estimator, $\mathcal{D}_{HS} =\langle \tr[(\rho_0 - \bar{\rho})^2]\rangle$  \cite{hradil02}, can be shown equal to the  total uncertainty in the Bloch vector components, $\mathcal{D}_{HS} = \sum_\alpha\langle (\Delta r_\alpha)^2 \rangle$. The Cramer-Rao inequlaity, $\langle (\Delta r_\alpha)^2 \rangle \ge \left[ \mbf{F}^{-1} \right]_{\alpha \alpha}$, relates these uncertainties to the   the Fisher information matrix, $\mbf{F}$, associated with the conditional probability distribution, Eq. (\ref{pdf_uniform_prior}), and thus $\mathcal{D}_{HS} \ge \tr[\mbf{F}^{-1}]$.  Since our probability distribution is a multivariate Gaussian regardless of the state, in the limit of negligible backaction, we saturate this bound.  In that case, the Fisher information matrix equals the inverse of the covariance matrix, $\mbf{F} = \mbf{C}^{-1}$, in units of $N^2/\sigma^2, $  {  where  $\mbf{C}^{-1}=\mathcal{\tilde{O}}^T \mathcal{\tilde{O}} ,$ and $\mathcal{\tilde{O}}_{n\alpha}= \tr [\mathcal{O}_n E_\alpha]$} \cite{madhok2014information}.  Thus, a metric for the total information gained in tomography is the inverse of this uncertainty,
\begin{equation}
\mathcal{J} = \frac{1}{\tr[\mbf{C}]}  
\end{equation}
which measures the total Fisher information.

In Fig. \ref{fig:FIrnd} we plot $\mathcal{J}$ as a function of time, generated by repeated application of the kicked top dynamics described above. We see a  close correlation between the level of chaos and the information gain in tomography for random states. Since the inverse covariance matrix is never full rank in this protocol, we regularize $\mbf{C}^{-1}$ by adding to it a small fraction of the identity matrix (see  e.g., \cite{boyd2004convex}). For pure states,  the average Hilbert-Schmidt distance $\mathcal{D}_{HS} =1/\mathcal{J}= 1-\langle \tr\bar{\rho}^2\rangle -2 \langle \mathcal{F} \rangle$~\cite{hradil02}. A correlation between chaos in the dynamics and the information gain as seen in the average fidelity (Fig. \ref{fig:fid1b}) implies that the Fisher information shows the behavior.

\begin{figure}[htbp]
 \centering
 \includegraphics[width=8.5cm, height=6.1cm]{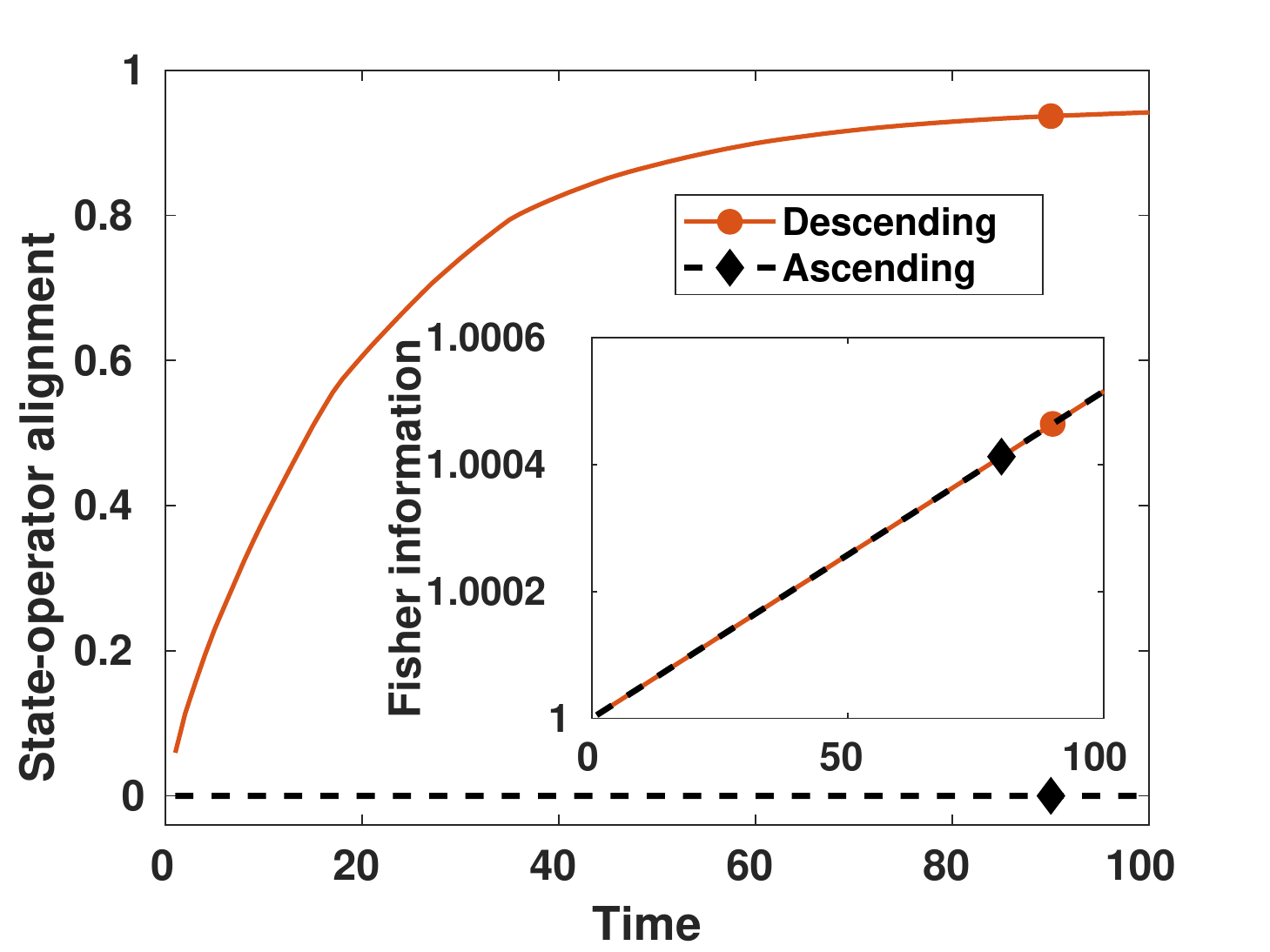} 
  
 \caption{{State-operator alignment  and Fisher information (the inset figure) for ordered $\{E_{\alpha}\}$ as a function of time. The solid line indicates the behaviour for the operators in descending order and the dotted line is for ascending order (i.e. that corresponds to the Bloch vector components, $r_{\alpha} = \tr[\rho_0 E_{\alpha} ]$  in the descending and ascending order of magnitude)}.}
 \label{fig:Fisov}
\end{figure}

Based on how much the dynamics generate Fisher information, the above analysis explains the reconstruction fidelity and its correlation with chaos for random states. However, the fact that Fisher information cannot capture all aspects of the problem can be easily seen by calculating it for the case 
discussed in Fig. \ref{fig:gell}.{ Since the Fisher information is independent of the order in which  $E_\alpha$'s are measured, it gives no information about the reconstruction procedure as shown in Fig. \ref{fig:Fisov}.} Therefore, we need to re-look at Eq. (\ref{pdf_uniform_prior}) and the prior information captured by the second term of Eq. (\ref{tom_eq}), $p(\rho_0|\mathcal{L}, \mathcal{M})$. 

As we have discussed, $p(\rho_0|\mathcal{L}, \mathcal{M})$ is the Bayesian estimate of the density matrix parameters at a particular time in the estimation process based on the information obtained. This is independent of the shot noise and depends on the nature of the observables measured  and the dynamics employed to generate these operators (choice of unitary). Thus, combining Eq. (\ref{pdf_uniform_prior}) and Eq. (\ref{tom_eq}), we get
\begin{widetext}
\begin{equation} 
\begin{split}
p({\bf \rho_0|M, \mathcal{L}, \mathcal{M}}) &\varpropto \mathrm{exp}\ \Big\{-\frac {N^2}{2\sigma^2}\sum_{i}[M_i-\sum_{\alpha}\mathcal{O}_{i\alpha}r_\alpha]^2\Big\}\      p(\rho_0|\mathcal{L}, \mathcal{M})  \\
&\varpropto  \mathrm{exp}\ \Big\{-{\frac {N^2}{2\sigma^2}\sum_{\alpha,\beta}({\bf r-r_{ML}})_\alpha\ C^{-1}_{\alpha\beta}\ ({\bf r-r_{ML}})_\beta\Big\}}\ p(\rho_0|\mathcal{L}, \mathcal{M}) 
\end{split}
\end{equation}
\end{widetext}

In the limit of zero shot noise, the errors due to the first term are zero and we may purely focus on 
the conditional probability distribution, $p(\rho_0|\mathcal{L}, \mathcal{M})$. In terms of the observables in continuous measurement tomography,  one can express $p(\rho_0|\mathcal{L}, \mathcal{M}) = p( \bf r| \mathcal{O}_1, \mathcal{O}_2, ..., \mathcal{O}_n )$,  giving the conditional probability of the density matrix parameters $\bf r$ till the time step $n$. For example, consider  the measurement operator at the first $k$ time steps are the ordered set \{$E_1$, $E_2$,...,$E_k$\}, giving precise information about Bloch vector components \{$r_1$, $r_2$, ..., $r_k$\}. The conditional probability distribution at time $k$ is,

{
\begin{widetext}
\begin{equation}
\begin{split}
p(  \textbf{r}| E_1,E_2, ..., E_k ) = \delta(r_1-\tr[E_1 \rho_0])\ \delta(r_2-\tr[E_2 \rho_0])\ ...\ \delta(r_k - \tr[E_k \rho_0])\ \delta \big(\sum^{d^2-1}_{\alpha \neq 1, 2,...k}r_{\alpha}^2 = 1 - 1/d - r_{1}^2 -  r_{2}^2 ...- r_{k}^2 \big).
\end{split}
\end{equation}
\end{widetext}
}

Each noiseless measurement above gives us complete information in one of the orthogonal directions. For example, at the first time step,
\begin{equation}
 p(\textbf{r}| E_1) =  \delta(r_1-\tr[E_1 \rho_0])\ \delta \big(\sum^{d^2-1}_{\alpha \neq 1}r_{\alpha}^2 = 1 - 1/d - r_{1}^2 \big).
\end{equation}
 Hence, once $r_1$ is determined, 
the rest of the $d^2 - 2 $ Bloch vector components are constrained to lie on a surface given by the equation $\sum^{d^2-1}_{\alpha \neq 1}r_{\alpha}^2 = 1 - 1/d - r_{1}^2 .$  
The state estimation procedure under incomplete information shall pick a state
consistent with $r_1$ as determined by the first measurement and the remaining Bloch vector components from a point on this surface. Therefore, qualitatively speaking, the average fidelity of the estimated state is correlated with the area of this surface. This area depends on the magnitude of $r_1$ that appears in the scaling factor mentioned above. Hence the order of measuring operators $\{E_{\alpha}\}$  that corresponds to the Bloch vector components \{$r_\alpha$\} in the descending order of magnitude gives the maximum rate of information gain as shown in Fig. \ref{fig:gell}. After $k$ time steps, the error is proportional to the area of the surface consistent with the equation $1 - 1/d - r_{i}^2 -  r_{j}^2- ... - r_{k}^2$. This area, quantifying the average error, shrinks with each measurement. The shrinkage rate of this error area for spin coherent states is more when the dynamics is regular. On the other hand, for random states, chaotic dynamics reveals more information about the initial condition as discussed above~\cite{madhok2014information}.

{To see it in another way, consider the fidelity between the actual and reconstructed state.} The fidelity $\mathcal{F}=\bra{\psi_0}\bar{\rho}\ket{\psi_0}$, combined with Eq. (\ref{dens_matrix}) for expressing both ${\rho_0}$ and $\bar{\rho}$, is

\begin{equation}
\mathcal{F} = 1/d+\Sigma^{d^2-1}_{\alpha=1}\ \bar{r}_\alpha {r}_\alpha
\end{equation}

As one makes measurements, ${E_{1}}$, ${E_{2}}$, ..., ${E_{k}}$ and gets information about the corresponding Bloch vector components (with absolute certainty in the case of zero noise for example), one can express the fidelity as 

\begin{equation}
\mathcal{F} = 1/d+\Sigma^{k}_{i=1}\  {r}_i^2 + \Sigma^{d^2-1}_{\alpha \neq 1, 2,...k}\ \bar{r}_\alpha {r}_\alpha 
\end{equation}

The term $1/d+ \Sigma^{k}_{i=1}\  {r}_i^2$ puts a lower bound on the fidelity obtained after $k$
measurements and, therefore, the rate of information gain in tomography is intimately tied with the extent of alignment between the measurement operators and the density matrix.

\begin{figure}[htbp]
 \centering
 \includegraphics[width=8.5cm, height=6.1cm]{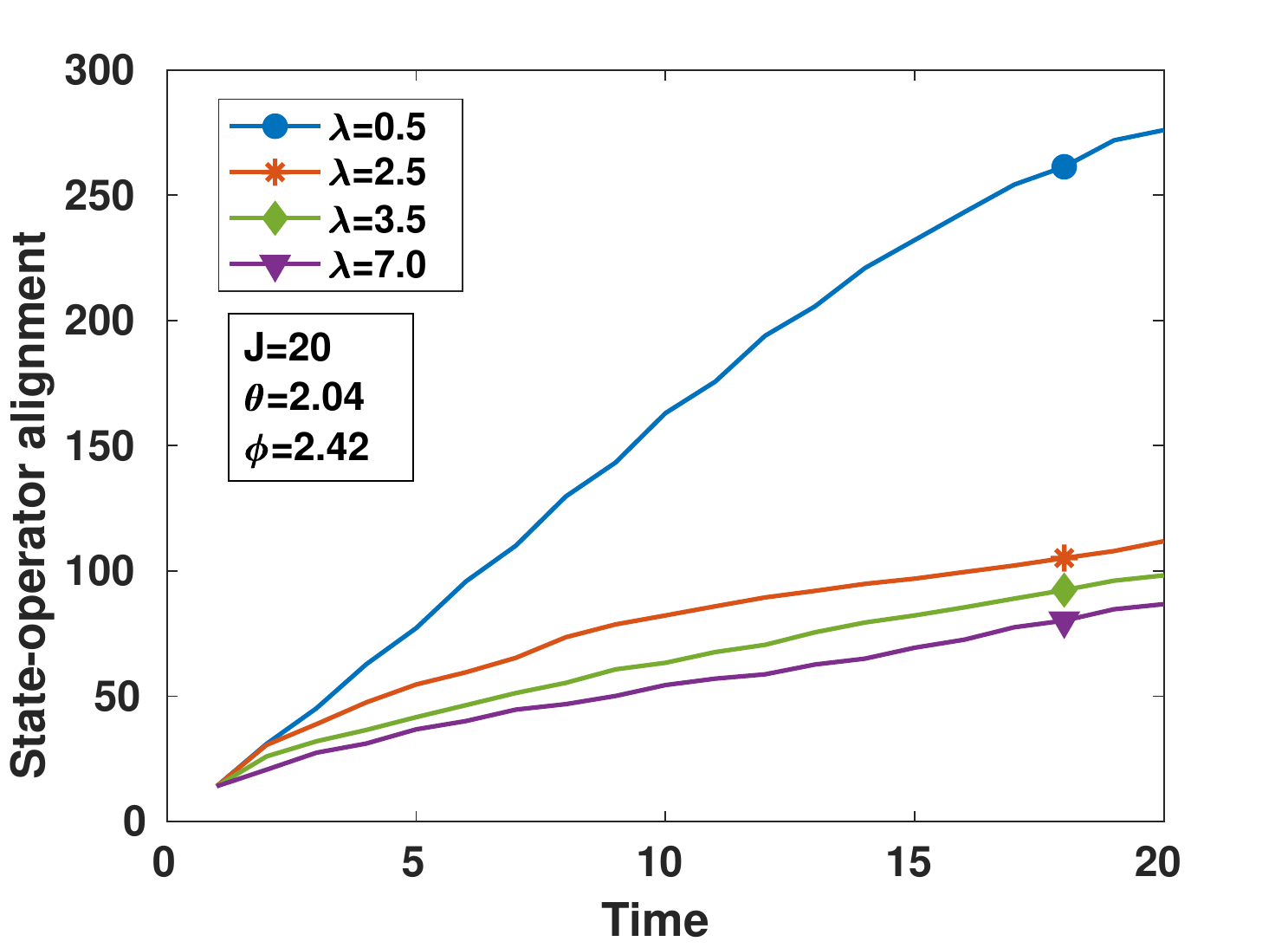} 
  
 \caption{State-operator alignment as a function of time for different degrees of chaos.}
 \label{fig:soalign}
\end{figure}

{The foregoing discussion helps us to understand how the ordering of operators facilitates the fidelity gain. Specifically, the overlap of the operators with the density matrix can be captured with the help of an ``alignment matrix"
\begin{equation}
\tilde{\mathcal{S}}=
\begin{pmatrix}
r_{1}\tilde{\mathcal{O}}_{11} & r_{2}\tilde{\mathcal{O}}_{12} & .. & .. & r_{d^2-1}\tilde{\mathcal{O}}_{1d^2-1}\\
r_{1}\tilde{\mathcal{O}}_{21} & r_{2}\tilde{\mathcal{O}}_{22} & .. & .. & r_{d^2-1}\tilde{\mathcal{O}}_{2d^2-1}\\
.. & .. & .. & .. & ..\\
.. & .. & .. & .. & ..\\
r_{1}\tilde{\mathcal{O}}_{n1} & r_{2}\tilde{\mathcal{O}}_{n2} & .. & .. & r_{d^2-1}\tilde{\mathcal{O}}_{nd^2-1}
\end{pmatrix} 
\end{equation}}
{where $\tilde{\mathcal{S}}_{n\alpha}=r_{\alpha} \tilde{\mathcal{O}}_{n\alpha} = r_{\alpha} \tr[\mathcal{O}_{n}E_{\alpha}]$, and $\mathcal{O}_n=U^{\dagger n}\mathcal{O}U^{n}$. }
{We quantify the extent of alignment of the time evolved operators with the state at a given time as $\tr[\mathcal{T}]$, where $\mathcal{T}=\tilde{\mathcal{S}}^{T}\tilde{\mathcal{S}}$.} State-operator alignment as shown in Fig. \ref{fig:Fisov}, explains the correlation between the information gain and the ordering of operators $\{E_\alpha\}$, while Fisher information is oblivious to that. {Figure \ref{fig:soalign} illustrates how the alignment of the operators with respect to the density matrix decreases with an increase in the degree of chaos, in agreement with the reconstruction rate of coherent states (Fig. \ref{fig:fid1a}). }

\begin{figure}[htbp]
  	\centering
  	\includegraphics[width=8.5cm, height=6.1cm]{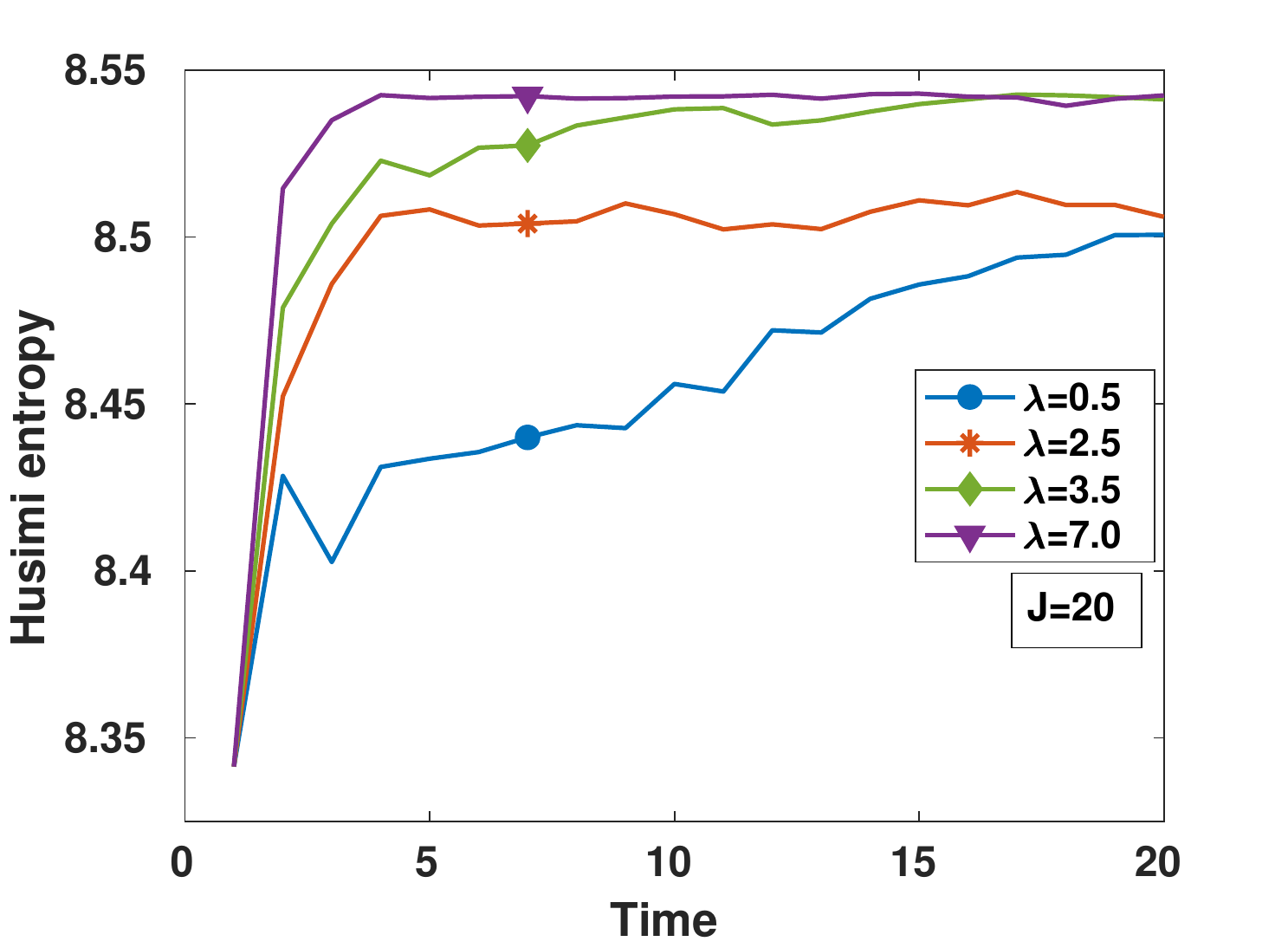} 
  	\caption{Husimi entropy of the operators evolved from the initial observable $\mathcal{O}=J_y$ as a function of time.}
  	\label{husimi}
  \end{figure}

{To understand the connection between state-operator alignment and the nature of the dynamics, } one can look at the localization of operators in the over-complete basis of spin coherent states. We notice that at a given time the operator becomes more delocalized as the chaos increases. This delocalization is captured by the Husimi entropy defined in Eq. (\ref{hs2}). The operator spreads more in the phase space as the chaoticity increases, which is apparent from  Fig. \ref{husimi}. The Husimi entropy increases and saturates at a higher value for a high value of chaoticity. A spin coherent state is localized in phase space and with the increase in chaos, the overlap of the state and the time evolved operator gets distributed in the phase space.  As the operator dynamics become more chaotic, more spin coherent states make up the operator, and the amount of information one gains about a particular state of interest is low.  Thus, the reconstruction of localized spin coherent states becomes difficult as the chaos in the dynamics shoots up. This behavior is also true for phase space averaged reconstruction fidelity of spin coherent states. 

  In contrast, the operator delocalization in phase space is positively correlated with the fidelity gain for random states. In this case, the most optimal measurement is the one that evenly measures all possible directions in the operator space and hence explains the positive correlation of information gain with the degree of operator spread in phase space. Interpreting this way, Tomography and information gain give us an operational interpretation of operator spreading and scrambling of information which is  being vigorously pursued through the study of out-of-time-ordered correlators and tripartite mutual information \cite{maldacena2016bound,swingle2016measuring,
  	hashimoto2017out,kukuljan2017weak,swingle2018unscrambling,wang2021quantum,sreeram2021out,varikuti2022out,seshadri2018tripartite}. This is in close resemblance with the classical Kolmogorov-Sinai (KS) entropy 
  which relates the increasingly fine-grained knowledge about the initial conditions as one monitors a chaotic trajectory \cite{sinai1959notion, caves1997unpredictability}.

\section{Discussion}

In this work, we have given a complete picture of the role of chaos in information gain in order to perform tomography via weak continuous measurements. 
Remarkably, the reconstruction rate of spin coherent states decreases with the increase in chaos, in contrast to the behavior of random quantum states. The fact is that the spin coherent states are localized in the phase space as a Gaussian wave packet with a minimal spread, unlike the random states, which are spread all over the phase space.{ The Fisher information serves as a suitable quantifier of information gain for random states where we consider only the dynamics. However, Fisher information does not reveal the behavior of decrease in the reconstruction rate of spin coherent states with an increase in chaos.  Thus, we include the prior knowledge and define a measure called state-operator alignment, which explains the decline in the fidelity rate as the dynamics become chaotic.
Furthermore, we show that the ordering of operators also plays a role in the reconstruction rate.} The angular momentum operators and the spin coherent states get delocalized in the phase space as we evolve them with chaotic dynamics. We see that the degree of delocalization of the operators increases with chaos. Hence, the information gain in the measurement decreases, making the reconstruction of spin coherent states more difficult.

Quantum tomography and quantum control are two sides of the same coin. Generating an informationally complete record requires sufficient non-integrability in the dynamics. This is the very same resource that drives a fiducial state to a target state. Therefore, an interesting consequence of our work is the quantum control of well-localized states, like the coherent states, using regular quantum dynamics. For example, one can accomplish quantum control by taking Gaussian states to Gaussian states with pure rotations. One would need chaotic quantum maps to take initial coherent states to target states that are random in nature.

Though quantum systems show no sensitivity to initial conditions, due to unitarity of evolution, they do show sensitivity to parameters in the Hamiltonian \cite{peres1984stability}. This leads to an interesting question for quantum tomography and, more generally, quantum simulations. Under what conditions are the system dynamics sensitive to perturbations, and how does this affect our ability to perform quantum tomography? Can quantum tomography say something about the notion of sensitivity to perturbations in system dynamics in quantum systems? In particular, one may ask, how do the effects of perturbations manifest in the reconstruction algorithm, and how are they affected by the chaoticity of the system?

Lastly, the connections between information gain, quantum chaos, and the spreading of operators are an exciting avenue providing an operational interpretation to operator scrambling, which is more popularly captured by out-of-time-ordered correlators (OTOCs).{ The information gain in tomography quantifies the amount of new information added as one follows the trajectory of operators generated by the dynamics in the Heisenberg picture. However, the OTOC is the quantum analog of divergence of two trajectories which is captured by Lyapunov exponents in the classical picture and operator incompatibility in the quantum counterpart \cite{larkin1969quasiclassical, maldacena2016bound,swingle2018unscrambling}. }Therefore, a natural direction is to connect the information gain in tomography to the Lyapunov exponents, thereby unifying the connections between information gain, scrambling, and chaos and connecting it to an actual physical process. We hope our work paves the way for future studies in this direction.

\section{Acknowledgements}
We are grateful to Arul Lakshminarayan for useful discussions.
This work was supported in part by grant number SRG/2019/001094/PMS from SERB and MHRD/DST grants SB20210807PHMHRD008128, SB20210854EEMHRD008074, DST/ICPS/QusT/Theme-3/2019/Q69 and New faculty Seed Grant from IIT Madras.

\bibliographystyle{unsrt}
\bibliography{cst}
\onecolumngrid
\end{document}